\documentclass{ws-procs975x65}
\usepackage{multirow}

\def\beq{\begin{equation}}
\def\eeq{\end{equation}}

\begin{document}

\title{(Not so) pure Lovelock Kasner metrics}




\author{Xi\'an O. Camanho }
\address{Max-Planck-Institut f\"ur Gravitationsphysik, Albert-Einstein-Institut, 14476 Golm, Germany}
\author{Naresh Dadhich}
\address{Centre for Theoretical Physics, Jamia Millia Islamia, New Delhi 110 025, India\\  Inter-University Centre for Astronomy \& Astrophysics, Post Bag 4 Pune 411 007, India\\
and Astrophysics and Cosmology Unit, School of Mathematics, University of KwaZulu Natal, Durban, South Africa. }
\author{Alfred Molina}
\address{Departament de F\'{\i}sica Fonamental, Institut de Ci\`encies del Cosmos. Universitat de Barcelona, Spain}
\date{\today}

\begin{abstract}
The gravitational interaction is expected to be modified for very short distances. This is particularly important in situations in which the curvature of spacetime is large in general, such as close to the initial cosmological singularity. The gravitational dynamics is then captured by the higher curvature terms in the action, making it difficult to reliably extrapolate any prediction of general relativity. In this note we review pure Lovelock equations for Kasner-type metrics. These equations correspond to a single $N$th order Lovelock term in the action in $d=2N+1,\,2N+2$ dimensions, and they capture the relevant gravitational dynamics when aproaching the big-bang singularity within the Lovelock family of theories. 
These are classified in several isotropy types. Some of these families correspond to {\it degenerate} classes of solutions, such that their dynamics is not completely determined by the equations of pure Lovelock gravity. Instead, these Kasner solutions become sensitive to the subleading terms in the Lovelock series. 

\end{abstract}


\bodymatter


\section{Introduction}

The gravitational dynamics close to a generic cosmological singularity has been extensively studied over the years.  The BKL conjecture (see Ref.~\refcite{Heinzle2009} for a recent discussion) proposes that the Universe close to the initial singularity becomes local, oscillatory and vacuum dominated,\cite{Belinskii2006} thus effectively undergoing a series of oscillations, transitions between Kasner epochs where the expanding and contracting directions exchange their roles. A precise realization of this behavior corresponds to the {\it Mixmaster universe} (see for instance Ref.~\refcite{Cornish1997}) of four dimensional general relativity.  

Even though classical gravity in four dimensions is well described by the Einstein-Hilbert action, it is also known that the interaction needs to be modified at high energies. Most approaches to a quantum theory of gravity involve higher order corrections in the curvature to the classical action. They appear in the context of Wilsonian approaches or as next-to-leading orders in the effective action of string theory. Most of these theories are problematic though, due to their higher derivative nature, and explicit analytic results are hard to extract. Besides, higher curvature interactions are particularly relevant when considering higher dimensional scenarios as we discuss below.  

We will restrict our attention to a very particular class of higher curvature theories that represent the most natural extension of general relativity to higher dimensions. Lovelock gravity is the most general gravitational theory yielding second order field equations,\cite{Lovelock1971} thus avoiding higher derivative ghosts.\cite{Zumino1986} Although  corrections appearing in the aforementioned frameworks are not generally of the Lovelock type, this family of theories presents many compelling features.  
They capture a wealth of characteristic higher curvature effects (existence of new branches of solutions corresponding to vacua with different effective cosmological constants, new instabilities, causality violation, etc.) in an analytically controllable setting,\cite{Camanho2015} 
providing a suitable playground to contrast our ideas about gravity in a much broader setup. The only drawback is that the first non-trivial higher order Lovelock  terms only appear above four dimensions. 


In this note we focus on Kasner-type metrics in (pure) Lovelock gravity. These solutions present a very simple form describing the leading order behavior close to a spacelike singularity such as the {\it Big-Bang} in standard cosmological scenarios. In our higher dimensional setting, the early time dynamics of a generic Lovelock theory will be given by the highest order term, thus pure Lovelock gravity, a single $N$th order term in odd $d=2N+1$ or even $d=2N+2$ dimensions.
 

A complete classification of Kasner type metrics has been obtained recently (see Table \ref{classes} in section 3 below).\cite{LLkasner} We will emphasize here the {\it degenerate} nature of some of the families of solutions identified. In those cases the dynamics is not completely determined by the pure Lovelock term and lower curvature corrections play an essential role as well. In particular, for a generic Lovelock theory for which all possible Lovelock terms are present in the action, the only non-trivial Kasner solutions correspond to those of type (b.2) in our classification, only present in the even dimensional case. More general spaces can be obtained when some of the terms in the Lovelock series are absent.

\section{Lovelock gravity}

The Lovelock gravity action and the corresponding field equations are given by a sum of homogeneous polynomial terms in the Riemann curvature, most simply written in terms of differential forms,
\begin{eqnarray}
\mathcal{L}&=&\sum_{k=1}^Nc_k \mathcal{L}_{(k)}=\sum_{k=1}^N\frac{c_k}{d-2k}\,\epsilon_{a_1 \cdots a_d}\,R^{a_1a_2} \cdots \wedge R^{a_{2k-1}a_{2k}}\wedge e^{a_{2k+1}} \cdots \wedge e^{a_d} ~,\\
G_{\ c}^{b}&=& \sum_{k=1}^N c_k  G_{(k)\ c}^{\quad b} = \sum_{k=1}^N \, c_k\,\epsilon_{a_1 \cdots a_{d-1} c}\,R^{a_1a_2} \cdots \wedge R^{a_{2k-1}a_{2k}}\wedge e^{a_{2k+1}} \cdots \wedge e^{a_{d-1}}\wedge e^b ~. \nonumber
\label{einstein}
\end{eqnarray}
Each term has an associated dimensionful coupling $c_k$, with length dimension $L^{2(k-1)}$ relative to the Einstein-Hilbert action. In this language the metric is parametrized by the vielbein $e^a$, whereas torsion and curvature forms are introduced via Cartan's structure equations,
\begin{eqnarray}
T^a &=& De^a=de^a+\omega^a_{\ b}\wedge e^b ~,\label{streq}\\
R^a_{\ b}&=&d\omega^a_{\ b}+\omega^a_{\ c}\wedge \omega^c_{\ b} ~.
\end{eqnarray}
In addition to the usual exterior operator $d$, we have introduced a covariant exterior derivative, $D$, and the corresponding connection 1-form $\omega^a_{\ b}$. In order to make contact with the usual tensorial formulation, one imposes that the torsion is zero and solves (\ref{streq})
 for the spin connection in terms of the vielbein. The equation of motion above comes from the variation of the action with respect to the vielbein only, leaving the spin connection unchanged. One may consider a second equation as arising from the independent variation of the latter form but this is automatically zero due to the torsionless condition. 

\section{Kasner metrics}

Kasner geometries represent the simplest instance of  homogeneous but anisotropic spacetimes. They have been widely used in the analysis and classification of cosmological singularities in the context of general relativity.\cite{Belinskii2006} These type of metrics were considered long ago by Deruelle  in the Gauss-Bonnet case.\cite{Deruelle1989} For a more general account of anisotropic models in Lovelock theories one may refer to Ref.~\refcite{Kitaura1991} and \refcite{Kitaura1993a}.

The form of the Kasner metric,
\begin{equation}
ds^2= \sum_{i=1}^n t^{2 p_i} dx_i^2 -dt^2~,
\label{Kasner}
\end{equation}
implies that the Riemann curvature grows as $t^{-2}$ as we go backwards in time, the leading contribution to the field equations coming from the highest order term in the action. We then restrict our analysis to pure Lovelock gravity in odd, $d=2N+1$ or even, $d=2N+2$, dimensions, even though, as we will see later, the subleading terms may also play an important role in some cases.

Under this conditions, the pure Lovelock field equation in vacuum reduce to a series of algebraic equations for the Kasner exponents, $p_i$. The isotropy conditions $G_{ii}-G_{jj}=0$ can be written as products of several factors, in such a way that, depending on which factors vanish, we can classify our solutions in several different isotropy types. These are summarized in the table below along with the isotropic perfect fluid and vacuum solutions. 

The number of free parameters for the vacuum families varies depending on the particular isotropy class. Both in odd and even dimensions we have two classes with $2N-2$ parameters each, whereas in even dimensions we have an extra family with one more free parameter. Moreover, one can easily see that type (b.1) metrics in even dimensions can be obtained simply as a limit of those of type (b.2), when two of the exponents are taken to be very small. 
It is also clear that, in three and four dimensions, type (c) vacuum solutions correspond to the trivial flat solution while we have non-trivial vacuum solutions in that class in dimensions above four. 

\begin{table}[ht]
\tbl{Classification of isotropic and vacuum solutions of $N$th order pure Lovelock gravity.  ${(\#_1)}$ $\mathcal{R}=0$ as well for the subset of type {\bf(b)} ({\bf(b.1)} in $d=2N+2$) with all non-zero exponents equal to one, $p_{i\neq1,2}=1$. ${(\#_2)}$ For {\it flat Kasner} (naively type {\bf(c)}) we also have $R=0$. For $N=2$, except for these {\it exceptional} cases, all the tensors that are not in the table are non-vanishing. }
{\begin{tabular}{c|c||c|c||c}
type & $d$ & isotropy cond. & vacuum & vac. curvature\\
\hline\hline
{\bf (a)} & any $d$ & $p_i=p \ ,\quad \forall i=1,2\ldots d-1$ & $p=0$ & $R=\mathcal{R}=\mathbb{R}=0$\\
\hline
{\bf (b)} & $2N+1$ &  & $p_1=0$ & $\mathbb{R}=0~$ $^{(\#_1)}$\\
\hskip.4in {\bf --b.1--}\, & $2N+2$  & $\sum_{i=1}^{d-1}p_i=2N-1$ & $p_1=p_2=0$ & $\mathbb{R}=0~$ $^{(\#_1)}$\\
\hskip.4in {\bf --b.2--}\, & $2N+2$  &  & $\sum_{i=1}^{d-1}\frac1{p_i}=0$ & $ R,\mathcal{R},\mathbb{R}\neq 0$\\
\hline
{\bf (c)} & $2N+1$ & $p_1=p_2=0$ & \multirow{2}{*}{\it all} & \multirow{2}{*}{$\mathcal{R}=\mathbb{R}=0~$ $^{(\#_2)}$}\\
& $2N+2$ & $p_1=p_2=p_3=0$ & & \\
\hline
{\bf (d)} & \multirow{2}{*}{$2N+2$} & $p_i=p_{1,2}$ with multiplicities $n_{1,2}$ & \multirow{2}{*}{\it none} & \\
&  & $\frac{n_1-1}{p_1}+\frac{n_2-1}{p_2}=0~, \quad n_1+n_2=2N+1$ & &
\end{tabular}}
\label{classes}
\end{table}


 The most important difference among the different vacuum families is the number of flat directions ($p_i=0$) in the corresponding geometries. Remarkably, this feature endows the different classes of solutions with discerning geometric properties. For instance, one may define two higher curvature 4th rank tensors, $\mathcal{R}^{(N)}_{abcd}$ and $\mathbb{R}^{(N)}_{abcd}$ (see Ref.~\refcite{LLkasner} and \refcite{xd} for the precise definitions of these tensors and their properties), besides the usual Riemann curvature, and use them to assign to each geometry its isotropy type. This may be useful when analysing asymptotic solutions that are not directly given on their canonical form. In table \ref{classes}, the reference to the curvature order and the indices of these tensors have been eliminated for simplicity.
 
Among all families of solutions, type (b.2) metrics are also the most generic in even dimensions. The solution space of this family has a very interesting geometric structure and symmetries, besides the obvious permutations of exponents, that may be useful when considering more complicated analyses.\cite{LLkasner} In the next section we will focus on metrics with flat directions, {\it i.e.} all families except type (b.2). As we will see all these solutions are {\it degenerate}, in the sense that their dynamics is not completely fixed by the pure Lovelock equations of motion. The predictability of the theory is lost, unless we are ready to give up the usual geometric interpretation for these metrics. 

\section{Degenerate solutions}

The higher order nature of the (pure) Lovelock lagrangian makes the dynamical content of the theory very non-trivial, with different sectors of solutions and different numbers of degrees of freedom on each sector.\cite{Zanelli2000a, Miscovic2005} The simplest example of this behavior happens for maximally symmetric spaces. These have a very simple form of the curvature 2-form, $R^{ab}=\Lambda\, e^a\wedge e^b$. For these spaces to be a solution of our Lovelock theory of choice, the {\it effective} cosmological constant of this space, $\Lambda$, has to be a solution of the polynomial defined by the Lovelock couplings,
\begin{equation}
\Upsilon[\Lambda]=\sum_{k=1}^K c_k \Lambda^k=c_K\prod_{i=1}^K (\Lambda-\Lambda_i)=0~.
\end{equation}
$\Lambda$ can be equal any real solution of the equation among the $K$ possible ones, $\Lambda_i$. 

Remarkably, the same polynomial $\Upsilon[\Lambda]$ controls many important aspects of the theory.\cite{Camanho2013,Camanho2015} For example, the equation for perturbations around any of these vacua is proportional to its first derivative. For a perturbation of a particular maximally symmetric solution (say $\Lambda_1$),  we may write the curvature as  $R^{ab}=\Lambda_1\, e^a\wedge e^b+\delta R^{ab}$ and get
\begin{equation}
G^b_{\ c}=\left(\sum_{k=1}^K k\,c_k\Lambda_1^{k-1} \right) {G}_{(1)}{}^b_{\ c}~,
\end{equation}
where ${G}_{(1)}{}^b_{\ c}$ is the usual Einstein ($k=1$) tensor. The prefactor can also be written as a function of the possible solutions of the polynomial simply as
\begin{equation}
\Upsilon'[\Lambda_1]=c_K\prod_{i\neq 1}(\Lambda_1-\Lambda_i)~.
\end{equation}
In this way it is clear that the equation for linearized perturbations vanishes exactly when the vacuum we are considering is {\it degenerate}, {\it i.e.} its effective cosmological constant $\Lambda_1$ is a multiple solution of the characteristic polynomial $\Upsilon[\Lambda]$. For a solution of multiplicity $n$ we have in general $\Upsilon[\Lambda_1]=\Upsilon'[\Lambda_1]=\Upsilon[\Lambda_1]''=\ldots =\Upsilon^{(n-1)}[\Lambda_1]=0$. Thus, linearized perturbations are completely undetermined by the equations of motions and it is sometimes said that the theory at the degenerate point {\it does not have linearized degrees of freedom}.\cite{Zanelli2000a} 

This could just be, of course, an artefact of the linearized approximation, however the same behavior is also observed  at the nonlinear level, for instance, in the existence of a new branch of spherically symmetric solutions for which the $g_{tt}$ component of the metric is a free function of at least some of the coordinates {\bf (including time)}.\cite{Bogdanos2009} For this new class of solutions the predictability of the theory is lost. For the same initial data we can consider an infinite number of solutions that differ simply in their time evolution after some later point. 

Something similar happens in pure Lovelock for some of the classes of Kasner solutions just described, in particular those with one or more flat ($p_i=0$) directions. In fact, pure Lovelock without a cosmological constant term in the action is {\it maximally} degenerate in the sense outlined above. We have a single vacuum, in this case just Minkowski, with maximal multiplicity $N$. However we will consider our equations not for maximally symmetric spaces but Kasner-type vacua, and the pattern of degeneracy is a bit more involved in that case.

For solutions on types (b) or (c) in odd $d=2N+1$ or (b.1) or (c) in even $d=2N+2$ dimensions, all the components of the Riemann tensor with any of the indices along any of the flat directions will yield zero. Moreover, for type (b) metrics, the number of flat directions is precisely the same as the number of vielbeine in the action or the equations of motion. This means that, keeping the flat part of the metric untouched, the equations of motion for a generic metric of the form
\begin{equation}
ds^2=\hat{g}_{\mu\nu}(y)\,dy^\mu dy^\nu+\sum_{i=1}^{n}dx_i^2 ~,
\label{flatansatz}
\end{equation}
with $n=1$ or $n=2$ in odd or even dimensions respectively, reduce simply to
\begin{equation}
\hat{\mathcal{L}}_{(N)}=0~.
\label{typeb}
\end{equation}
The lower dimensional Lovelock density associated with $\hat{g}$ has to vanish. It is clear that this single equation will not be enough to completely determine the evolution of this $(d-n)$-dimensional metric. In particular, for a Kasner solution, any perturbation of $\hat{g}$ that is not of the form $\delta \hat{R}^{ab}=F(y)\,\hat{e}^a \wedge \hat{e}^b$ will be completely unconstrained. This of course carries on to the nonlinear regime. This is easy to see, as we may write many different tensorial equations of the form
\begin{equation}
\hat{\mathfrak{R}}_{ab}(\hat{g})=\mathcal{T}_{ab}(y)\qquad \text{with} \qquad \hat{\mathfrak{R}}^a_{\ a}= \hat{\mathcal{L}}_{(N)}~,
\end{equation}
such that the trace of the new equation reduces to (\ref{typeb}). Moreover, the tensor $\mathcal{T}_{ab}$ may have an arbitrary dependence on the $y$-coordinates as long as it is traceless. Each of these equations would in general yield a different metric for each different $\mathcal{T}_{ab}$. 

The situation is even more dramatic for type (c) metrics for which the number of flat directions $n$ is bigger than the number of vielbeine in the action. Thus, for a metric of the form (\ref{flatansatz}), the pure Lovelock field equations are identically zero, irrespective of the form of $\hat{g}$. The lower dimensional metric is completely free in that case. 

In the above cases the {\it degeneracy} of the solutions is lower than in the case of degenerate maximally symmetric spaces. We still get some equations for some subset of the perturbations around that particular spacetime, in particular for those that do not respect the ansatz (\ref{flatansatz}). It is important to point out that the perturbations that are not constrained by the equations are not just gauge degrees of freedom but physical ones, in the sense that they do not amount to a change of coordinates of the original metric. In some cases, it is possible to identify extra gauge symmetries of the theory at the degenerate sector,\cite{Zanelli2000a} in such a way that we can interpret the extra unconstrained modes as pure gauge as well. One may restore the predictability of the theory in this way, at the price of giving up   the usual geometric interpretation of our solutions. The new gauge symmetry group encompasses the usual diffeomorphism invariance within a larger group, meaning it identifies as {\it physically equivalent} metrics with different geometric properties. 

Of course, the above arguments only apply to an action that is {\it exactly} pure Lovelock, not when a solution of a more general Lovelock theory approaches (pure Lovelock) Kasner at early times. The previous degeneracy would then be lifted and the solution would have in general a well defined evolution. The dynamics of the components not determined by the pure Lovelock equations will instead be given by the next non-zero lower curvature order in the equations. This will have important consequences for our classification of Kasner solutions in more general Lovelock models and also for the stability of these spacetimes. 

\section{Lower curvature corrections}

Coming back to our ansatz (\ref{flatansatz}), it is easy to discuss how the rest of the Lovelock series affects our classification. The type (c) case is the simplest to analyze. In that case the $N$th order contribution to the Lovelock equations vanishes exactly and we are left with the first subleading term, $k<N$. The structure of the equations would then be,
\begin{eqnarray}
\hat{G}_{(k)}{}^b_{\ c}&=&0 ~,\\
\hat{\mathcal{L}}_{(k)}&=&0~,
\end{eqnarray}
for the odd dimensional ($\hat{d}=2(N-1)+1$) metric $\hat{g}$. The first line comes from the equations along the directions of $\hat{g}$, whereas the scalar constraint comes from the flat directions. There are no mixed field equations. The scalar constraint is actually proportional to the trace of the first line and can therefore be neglected. We are left with the usual $k$th order Lovelock equations for the reduced metric. This reduction ansatz is not consistent when considering more than one Lovelock term in general,\cite{Giribet2006} but we are only looking at the leading contribution here. In the particular case where the first subleading term is actually $k=N-1$, we end up with $(N-1)$th pure Lovelock in the critical (odd) dimension $\hat{d}$. For a Kasner ansatz, we can again classify solutions using our table above.  

For type (b) metrics (type (b.1) in even dimensions) we can play a similar game. The only difference is that in this case the contribution of the leading $N$th term does not vanish completely. The analog of the scalar constraint  will have an extra term proportional to $\hat{\mathcal{L}}_{N}$ in addition to $\hat{\mathcal{L}}_{k}$. Again, the latter piece is just proportional to the trace of the remaining equations but the former remains. We get the $k$th order Lovelock equations for the even ($\hat{d}=2N$) dimensional metric $\hat{g}$, supplemented by the $N$th order constraint, $\hat{\mathcal{L}}_{N}=0$. 

The lower dimensional metric seems to be overconstrained. In fact, this extra condition will be incompatible with the Kasner ansatz in many cases. For $k=N-1$ we end up again with $(N-1)$th order pure Lovelock  in the relevant even dimension, $\hat{d}=2(N-1)+2$. Our solutions can be classified again into several isotropy classes, but we only need to consider those of type (b.2). The remaining families have extra flat directions and fall in the type (c) scenario discussed above. Unfortunately one of the constraints of type (b.2) metrics is incompatible with the extra scalar equation. When, none of the exponents of $\hat{g}$ is zero, the $N$th order constraint amounts to 
\begin{equation}
\sum_i \hat{p}_i=2N-1
\end{equation}
that is incompatible with that same sum being equal to $2N-3$ as required by the type (b.2) condition on $\hat{g}$. The same would happen when the action consists only on the Einstein term and the highest order one in the Lovelock family. In that case the Ricci flat condition would imply that the sum is actually one. 

Therefore, the type (b) ansatz (type (b.1) in even dimensions) is incompatible with the presence of the $(N-1)$th order Lovelock term in the action. In this context, even dimensional type (b.1) metrics have to be understood only as a limit of those of type (b.2). 
In that case we are left with types (c) and (b.2), the latter just for even dimensionality. Then, for type (c) metrics we have to consider pure Lovelock equations on $\hat{g}$ that is odd dimensional and, in case the $N-2$ order term is also present, $\hat{g}$ has to be type (c) as well. When all of the Lovelock terms ($k>0$) are present in the action,  the iteration of the above process would imply that the only consistent Kasner solution is Minkowski.

\section{Discussion}

In this note we have extended our classification of pure Lovelock Kasner solutions of Ref.~\refcite{LLkasner} to more general Lovelock models. Except for the so-called type (b.2) spaces in even dimensions, Kasner metrics represent examples of degenerate solutions for pure Lovelock gravity, 
although the issue is solved by the inclusion of some lower curvature corrections ($k>0$). As we just saw, these corrections are incompatible with some of the initial isotropy types identified, namely types (b) or (b.1) in odd and even dimensions respectively, and lead to an iterative classification scheme.

The only nontrivial solutions would correspond in that case to type (b.2) metrics, unless some of the terms in the Lovelock series are absent. For instance, in case the Einstein-Hilbert term is the only subleading correction, the dynamics of $\hat{g}$, the lower dimensional metric in (\ref{flatansatz}), would reduce to the usual general relativistic one. In this more involved situation, a general identification of the different types of solutions may be achieved using tensors analogous to the pure Lovelock case but lower order in curvature: $\mathcal{R}^{(k)}_{abcd}$, $\mathbb{R}^{(k)}_{abcd}$ and possibly their higher rank analogs.\cite{xd} Moreover, some parts of the analysis performed here can be easily extended to more general product spaces of the form $\mathcal{M}=\mathcal{M}_1\times\mathcal{M}_2$, with arbitrary metric on each factor.\cite{products} 
 
Our classification of Kasner solutions in Lovelock models  represents a necessary first step in the general analysis of cosmological singularities in higher curvature  gravitational theories. We expect to report more on this soon.

\section*{Acknowledgments}


ND would like to thank Hermann Nicolai for a visit to the Albert Einstein Institute which facilitated this work.

\end{document}